\input harvmac

%
%
%
\def\ehp{ {\hat{e}}\, ' } 
\def\col{\pmatrix{\xp \pm \xm \cr ~ \xph \pm \xmh \cr } } 
\def\Ch{\hat{C}} 
\def\kb{{\bf K}}
\def\rb{{\bf R}}
\def\bra#1{{\langle #1 |  }}

\def\rb{ \right]  }

\def\bar{\overline}
\def\hat{\widehat}
\def\*{\star}
\def\[{\left[}
\def\]{\right]}
\def\({\left(}		
\def\){\right)}

%
%
\def\zb{{\bar{z} }}
\def\frac#1#2{{#1 \over #2}}
\def\inv#1{{1 \over #1}}
\def\half{{1 \over 2}}
\def\d{\partial}

\def\vev#1{\langle #1 \rangle}
\def\ket#1{ | #1 \rangle}
\def\rvac{\hbox{$\vert 0\rangle$}}
\def\lvac{\hbox{$\langle 0 \vert $}}
\def\2pi{\hbox{$2\pi i$}}

\def\dsl{\raise.15ex\hbox{/}\kern-.57em\partial}
\def\Dsl{\,\raise.15ex\hbox{/}\mkern-.13.5mu D}
%
%
\def\th{\theta}

\def\al{\alpha}
\def\ep{\epsilon}

\def\sig{\sigma}	\def\Sig{\Sigma}

%
%

	\def\CN{{\cal N}}

\def\rvac{\hbox{$\vert 0\rangle$}}
\def\lvac{\hbox{$\langle 0 \vert $}}

\def\2pi{\hbox{$2\pi i$}}

\def\dsl{\raise.15ex\hbox{/}\kern-.57em\partial}
\def\Dsl{\,\raise.15ex\hbox{/}\mkern-.13.5mu D}
%
%
%
\font\numbers=cmss12
\font\upright=cmu10 scaled\magstep1
\def\stroke{\vrule height8pt width0.4pt depth-0.1pt}
\def\topfleck{\vrule height8pt width0.5pt depth-5.9pt}
\def\botfleck{\vrule height2pt width0.5pt depth0.1pt}
\def\Zmath{\vcenter{\hbox{\numbers\rlap{\rlap{Z}\kern
0.8pt\topfleck}\kern
2.2pt
                   \rlap Z\kern 6pt\botfleck\kern 1pt}}}
\def\Qmath{\vcenter{\hbox{\upright\rlap{\rlap{Q}\kern
                   3.8pt\stroke}\phantom{Q}}}}
\def\Nmath{\vcenter{\hbox{\upright\rlap{I}\kern 1.7pt N}}}
\def\Cmath{\vcenter{\hbox{\upright\rlap{\rlap{C}\kern
                   3.8pt\stroke}\phantom{C}}}}
\def\Rmath{\vcenter{\hbox{\upright\rlap{I}\kern 1.7pt R}}}
\def\Z{\ifmmode\Zmath\else$\Zmath$\fi}
\def\Q{\ifmmode\Qmath\else$\Qmath$\fi}
\def\N{\ifmmode\Nmath\else$\Nmath$\fi}
\def\C{\ifmmode\Cmath\else$\Cmath$\fi}
\def\R{\ifmmode\Rmath\else$\Rmath$\fi}


\def\euiN{{ \( \prod_{i=1}^N \exp (  -mz\> u_i - m \zb \> u_i^{-1} ) \) }} 
\def\eui{{ \( \prod_{i=1}^{2n}  \exp (  -mz\> u_i - m \zb \> u_i^{-1} ) \) }} 
\def\eu{{ \exp \( -\frac{m}{2} (zu + \zb u^{-1} ) \) }}
\def\ep{\varepsilon}
\def\Det{{\rm Det}}
\def\sg{SG_{ff} } 
\def\eh{{\hat{e}}} 

\Title{CLNS 94/1276, SPhT-94-021, hep-th/9402144}
{\vbox{\centerline{Differential Equations}
\centerline{ for Sine-Gordon Correlation Functions   }
\centerline{ at the Free Fermion Point
} 
}}

\bigskip
\bigskip

\centerline{Denis Bernard\foot{Member of CNRS.} }
\medskip
\centerline{Service de Physique Th\'eorique, CEN-Saclay\foot{Laboratoire
de la Direction des sciences de la mati\`ere du Commissariat \`a 
l'\'energie atomique.} } 
\centerline{F-91191 Gif sur Yvette, France}
\bigskip
\centerline{Andr\'e LeClair}
\medskip\centerline{Newman Laboratory}
\centerline{Cornell University}
\centerline{Ithaca, NY  14853}
\bigskip\bigskip

\vskip .3in

We demonstrate that for the sine-Gordon theory at the free
fermion point, the 2-point correlation functions of the
fields $\exp (i\al \Phi )$ for  $0< \al < 1$ can be
parameterized in terms of a solution to a sinh-Gordon-like
equation.  This result is derived by summing over intermediate
multiparticle states and using the form factors to express
this as a Fredholm determinant.  The proof of the
differential equations relies on a $\Zmath_2$ 
graded multiplication law satisfied by the integral
operators of the Fredholm determinant. Using this methodology,
we give a new proof of the differential equations which
govern the spin and disorder field correlators in the
Ising model.

\Date{2/97}
%
%
%
%
%
%
%
%
%
%
%
%
%
%
%
%

\newsec{Introduction}

It is well-known that for certain integrable quantum field
theories in 2 dimensions, non-trivial correlation functions
can be parameterized by solutions of non-linear partial
differential equations\foot{This manuscript
is a corrected version
of our work which has appeared in Nucl. Phys. B426 (1994) 534.
Previously, we had not found the most general solution to
the differential equations (3.37), but rather projected onto
a solution which is valid only for $\alpha = \alpha'$.
The only changes in this new version are at the end of section 3,
where the general solution for $\alpha \neq \alpha'$ are obtained.
We thank S. Lukyanov for drawing our attention to a possible error.
Recently, similar results were obtained by H. Widom using different
methods\ref\rwidom{H. Widom, {\it An Integral Operator Solution
to the Matrix Toda Equations}, solv-int 9702007.}. }.
The first result of this nature
was obtained by Wu et. al.\ref\rwu{T. T. Wu, B. M. McCoy,
C. A. Tracy, and E. Barouch, Phys. Rev. B13 (1976) 316.},
where it was shown that the 2 point correlation functions of spin 
and disorder fields can be parameterized in terms of a solution
to the Painlev\'e III equation.  Subsequently, this
result was obtained in a variety of ways\ref\rmjs{M. Sato, M. Jimbo,
and M. Miwa, Lecture Notes in Phys. vol. 126, 
Springer 1980, p. 429-491}\ref\rbb{O. Babelon and D. Bernard, Phys. Lett.
B288 (1992) 113.}. 
Miwa, Jimbo and Sato obtained the result starting from the braiding
relations of the spin fields with the free fermion fields. 
In \rbb, using the form factors, the sum over intermediate
multiparticle states was related to tau-functions satisfying
known differential equations.  The two latter derivations are
not unrelated, since in \rmjs\ the form-factors are implicit
in the expressions for the spin fields in terms of the fermions. 

Differential equations for correlation functions in other models
which possess free fermion points were derived by Its   
et. al.\ref\rkor{A. R. Its, A. G. Izergin, V. E. Korepin and 
N. A. Slavnov, Int. J. Mod. Phys. B4 (1990) 1003.}\ref\book{V. E.
Korepin, N. M. Bogoliubov, and A. G. Izergin, {\it Quantum
Inverse Scattering Method and Correlation Functions}, Cambridge
Univ. Press, 1993.}.  In this approach, the Bethe ansatz is
used to express the correlation function as a Fredholm determinant,
and general techniques were developed for obtaining differential
equations from this representation. 

In this paper, we derive differential equations that characterize
the non-trivial correlation functions of the sine-Gordon theory
at the free fermion point ($SG_{ff}$).  The method we use is a kind
of hybrid of the methods discussed above.  Namely, using the 
form factors of the fields, we express the sum over intermediate
multiparticle states as a Fredholm determinant. The type of 
kernels we obtain differ from the ones considered 
in \rkor\book. Nevertheless,
we find that the methods developed there can be
generalized to deal with the relativistic kernels which
generically arise from the form factor sum.  Indeed, using
this methodology, we also give a new proof of the differential
equations for the Ising theory.  In should be pointed out that 
Fredholm determinant expressions in the $SG_{ff}$ theory were
also obtained in \ref\rikt{H. Itoyama, V. E. Korepin, and 
H. B. Thacker, Mod. Phys. Lett B6 (1992) 1405.}, however 
the latter determinants have a different origin than the
form factor sum, namely they originate from the Bethe ansatz,
and do not coincide with our expressions. 

In \ref\smjb{M. Sato, T. Miwa and M. Jimbo, Publ. RIMS, Kyoto
Univ., 15 (1979) 871.}\ the correlation functions of fields
with non-trivial monodromy with charged free fermion fields
(as in eq. 2.14 below) were studied.  Since such fields are
essentially completely characterized by this monodromy, one 
can interpret the work 
 \smjb\ as dealing with the sine-Gordon theory,
and the results on differential equations presented below 
are implicit in \smjb.  The methods we use are  different
and perhaps more direct.  Generalization of the methods employed
in \smjb\ to the n-point functions is discussed in \ref\smjbc{M. Sato,
T. Miwa and M. Jimbo, Publ. RIMS, Kyoto Univ. 15 (1979) 577.}

The $SG_{ff}$ theory
is perhaps the simplest relativistic quantum field theory
where all of the integrability structures have been developed,
such as the Lax pair, classical and quantum monodromy matrix,
quantum inverse scattering, etc., and furthermore has 
non-trivial correlation functions of the basic fields on
which these structures are based. This is to be contrasted with the
situation in Ising, where even a lagrangian involving the 
spin fields is unavailable. For this reason, we believe 
our $SG_{ff}$ results may be instrumental for investigations
beyond the free fermion point. 

The result we derive is simply described. The sine-Gordon theory
is described by the Euclidean action
\eqn\action{
S = \inv{4\pi} \int d^2 z \( \d_z \Phi \d_\zb \Phi + \lambda 
\cos (\hat{\beta} \Phi ) \),  }
where 
$z = (t+ix)/2 $, $\zb = (t-ix)/2 $.  The free fermion point
occurs at $\hat{\beta} = 1$\ref\rcol{S. Coleman, Phys. Rev. 
D 11 (1975) 2088.}. Consider the correlation functions
\eqn\eIi{
\lvac : e^{i\al \Phi (z , \zb ) } : \> :e^{i \al ' \Phi (0) } : \rvac 
\equiv \exp \Sigma (z , \zb ), }
where 
$0< \al , \al' < 1$.  Since the bosonization relations
are $: \exp (\pm i \Phi ) : = \psi_\pm \bar{\psi}_\mp$, 
where $\psi_\pm , \bar{\psi}_\pm $ are the free massive
Dirac fermions, the fields $:\exp (i \al \Phi ):$ 
for $0<\al < 1$ are not simply expressed in terms of the
fermions, and thus have non-trivial correlation functions. 
We show that these correlation functions
can be parameterized by a solution $\varphi (r )$
of a  differential  equation
\eqn\result{\eqalign{
\( \d_r^2  + \inv{r} \d_r \)  \Sigma (r ) & =  \frac{m^2}{2} 
\( 1 - \cosh 2\varphi \) \cr
\( \d_r^2 + \inv{r} \d_r \)  \varphi &= \frac{m^2}{2} \sinh 2\varphi 
+ \frac{4 (\alpha - \alpha')^2 }{r^2} \tanh \varphi (1-\tanh^2 \varphi )
, \cr}}
where
$m$ is the mass of the free particles, and $r^2 = 4 z \zb$. 
We will also show that 
the result\result\ continues to be true if one replaces
the vacuua in  \eIi\ with multiparticle states. 
The equivalence of the differential equations \result\ 
and the Painlev\'e ones is well known. (See e.g. 
\ref\jimbo{M. Jimbo, Progr. Theor. Phys. 61 (1979) 359.}.) 

As discussed below, when $\al , \al' = \pm 1/2$, the 
$SG_{ff}$ correlation functions can be related to squares
of Ising correlation functions\ref\riz{J. B. Zuber and 
C. Itzykson, Phys. Rev. D 15 (1977) 2875.}\ref\rst{B.
Schroer and T. T. Truong, Nucl. Phys. B144 (1978) 80.}. 
Our results are obtained entirely within the $SG_{ff}$ theory
and do not use this connection with ${\rm Ising}^2$. 

\vfill\eject

\newsec{Correlation Functions as Fredholm Determinants} 

\bigskip
\noindent
2.1 {\it Form Factors} 

\medskip

The particle spectrum of the  $\sg$ theory consists of free solitons 
with topological $U(1)$ charge $\ep = \pm 1$, where   the solitons are 
identified as the free charged fermions.  The N-particle states
will be denoted as 
\eqn\eIIi{
| \th_1 , \cdots , \th_N \rangle_{\ep_1  \cdots  \ep_N} , } 
where 
$\th_i$ ($\ep_i$) is the rapidity ($U(1)$ charge) of the i-th particle, 
$E_i = m \cosh \th_i $, $P_i = m \sinh \th_i $. Inner products are
normalized as follows: 
\eqn\eIIii{
{}^{\ep'} \langle \th '| \th \rangle_\ep = 
\delta^{\ep '}_{\ep} ~ \delta(\th - \th' ) .}
One then has the following resolution of the identity:
\eqn\eIIiii{
1 = \sum_{N=0}^\infty \inv{N!} \sum_{\ep_1 \cdots \ep_N }  
\int_{-\infty}^\infty d\th_1 \cdots \int_{-\infty}^\infty 
d\th_N ~~ 
|\th_1  \cdots  \th_N \rangle_{\ep_1 \cdots \ep_N }      
{}^{\ep_N \cdots \ep_1 }
\langle \th_N \cdots \th_1 |  ~ . } 

Consider now the quantum fields $\sigma_\al$, 
\eqn\eIIiiib{
\sigma_\al (z , \zb ) \equiv  ~ : e^{i\al \Phi (z ,\zb ) } :  } 
where 
$\Phi$ is the SG field with the normalization implicit in the
action \action.  These fields are local, Lorentz spinless, 
and $U(1)$ neutral.  They have the hermiticity property: 
\eqn\eIIiv{
\( \sigma_\al \)^\dagger = \sigma_{-\al}  . } 
In terms of the $U(1)$ charged fermion fields $\psi_\pm , 
\bar{\psi}_\pm $, the bosonization relations are 
$: \exp (i\Phi ):  = \psi_+ \bar{\psi}_- $, 
$: \exp (-i\Phi ):  =  \bar{\psi}_+ \psi_-   . $ 
Thus, we limit ourselves to the range 
$0 < \al < 1 $.  Fields outside this range can be obtained
through operator products with fermion bilinears:
\eqn\eIIvi{
\sigma_{\al \pm 1 } \sim  
: \psi_\pm \bar{\psi}_\mp \> \sigma_\al :  } 

The form factors of these fields are defined as 
\eqn\eIIvii{
F_\al (\th_1 , \cdots , \th_N )_{\ep_1 \cdots \ep_N}  \equiv 
\lvac \sigma_\al (0) | \th_1  \cdots  \th_N \rangle_{\ep_1 \cdots \ep_N} .} 
Since the fields $\sigma_\al $ are $U(1)$ neutral, the form factors
are non-vanishing only for $U(1)$ neutral states. 
Explicit expressions for the form factors are as follows. 
For convenience define 
\eqn\eIIviii{
f_\al (\th_1 , \cdots , \th_{2n} ) 
= 
F_\al (\th_1 , \cdots , \th_{2n} )_{\ep_1 \cdots \ep_{2n}}
, ~~~~~~
\ep_i = - \ep_{i+n} = -1, ~ i=1,..,n .} 
It is also helpful to define the momentum space variable 
\eqn\eIIix{
u \equiv \exp (\th ) . } 
Then one has 
\eqn\eIIx{\eqalign{
f_\al (\th_1 \cdots \th_{2n} )
&= (c_\al m )^{\al^2} ~ (-1)^{n(n-1)/2 }
\(\frac{\sin (\pi\al )}{i\pi} \)^n
\( \prod_{i=1}^n (u_i)^{\half-\al}(u_{i+n})^{\half+\al}\)
\cr 
&~~~~~~~~~~\times 
\Delta (u_1 , \cdots , u_{2n} ) ,  \cr }}
where 
\eqn\eIIxi{
\Delta (u_1 , \cdots , u_{2n} ) =
\frac{\prod_{i<j \leq n } (u_i - u_j )
\prod_{n+1 \leq i < j } ( u_i - u_j ) }{
\prod_{r=1}^n \prod_{s=n+1}^{2n} (u_r + u_s )   . }} 

The factor $(c_\al m )^{\al^2}$ is included based on our knowledge
of the scaling dimension of the field $\sigma_\al$, which is the
same as in the massless limit.  The dimensionless constants $c_\al$ 
can be fixed from knowledge of the 1-point functions: 
\eqn\eIIxib{
\langle 0 | \sigma_\al (0) |0\rangle = (c_\al m)^{\al^2} . }
We henceforth set $c_\al m = 1 $. 

The form factors for other assignments of the charges 
$\ep_i$ are trivially related to $f_\al$ by the relation
\eqn\eIIxii{
F_\al (\th_1 , \cdots, \th_i , \th_{i+1} , \cdots ,\th_N )_{\ep_1 
\cdots \ep_i \ep_{i+1} \cdots \ep_N} 
= - 
F_\al (\th_1 , \cdots, \th_{i+1} , \th_i , \cdots ,\th_N )_{\ep_1 
\cdots \ep_{i+1} \ep_i \cdots \ep_N} 
. } 

The above form factors can be derived in a variety of ways. They
were first obtained implicitly by Schroer and Truong\rst\
by first expressing the SG field as a non-local fermion bilinear 
using $\d_z \Phi = \psi_+ \psi_- $, $\d_\zb \Phi = - \bar{\psi}_+ 
\bar{\psi}_- $, and then normal ordering the expressions 
$\exp (i\al \Phi )$ with respect to the fermions.  
For $\al = \pm 1/2$, these form factors were also obtained by
Smirnov using the bootstrap axioms\ref\rsmir{F. A. Smirnov,
{\it Form Factors in Completely Integrable Models of Quantum
Field Theory},  Advanced Series in Mathematical Physics 14, World
Scientific, 1992.}.  In 
appendix A, we give an independent derivation of the form factors
based on the monodromy relations: 
\eqn\eIIxiii{\eqalign{ 
\sigma_\al (x, t ) \> \psi_\pm (y,t) 
&= \psi_\pm (y,t) \> \sigma_\al (x,t) ~~~~~~~~~~~~~~~~ {\rm for}\ y<x  \cr
&= e^{\pm 2\pi i \al } ~ \psi_\pm (y,t) \> \sigma_\al (x,t) ~~~~~ 
~~{\rm for}\ y>x  \cr
}}
The latter derivation is an extension of techniques developed by 
Jimbo, Miwa, and Sato for the Ising theory\rmjs. 
Finally, we mention that the form factors can also be derived using
vertex operator techniques in radial quantization   
\ref\rlec{A. LeClair,
{\it Spectrum Generating Affine Lie Algebras in Massive Field
Theory}, CLNS 93/1220, hep-th/9305110, to appear in Nucl. Phys. B;
C. Efthimiou and A. LeClair, {\it Particle-Field Duality 
and Form Factors from Vertex Operators}, CLNS 93/1263, hep-th/9312121.}.

We now describe some of the main properties of the above
form factors.  The hermiticity relation \eIIiv\
necessarily implies 
\eqn\eIIxiv{
F_\al (\th_1 \cdots \th_N )_{\ep_1 \cdots \ep_N} 
= F_{-\al} (\th_N  \cdots \th_1 )_{\ep_N^* \cdots \ep_1^*} , } 
where 
$\ep^* = -\ep$.  The expression \eIIx\ is primarily characterized
by the residue property: 
\eqn\eIIxv{
2\pi i ~ f_\al (\th_1 \cdots \th_n , \th_n + i\pi + \gamma , 
\th_{n+2} \cdots \th_{2n} ) 
\sim 
\frac{(1-e^{2\pi i \al } )}{\gamma} 
f_\al (\th_1 \cdots \th_{n-1} , \th_{n+2} , \cdots \th_{2n} ) . }
The above relation is a generalization of the residue bootstrap
axiom for form factors.  As explained in \ref\rdouble{D. Bernard
and A. LeClair, Nucl. Phys. B399 (1993) 709.}, 
the additional phase $\exp (2\pi i \al )$ is a consequence of
the monodromy \eIIxiii. 

\bigskip
\noindent
2.2 {\it Fredholm Determinants} 
\medskip

Consider first the 2-point correlation functions. 
Inserting the resolution of the identity one obtains 
\eqn\eIIxvi{\eqalign{
\langle 0 | \sigma_\al (z , \zb ) \> \sigma_{\al'} (0) 
| 0 \rangle 
&= \sum_{N=0}^\infty \inv{N!} 
\sum_{\ep_1 , \cdots , \ep_N} 
\int d\th_1 \cdots d\th_N ~ 
\euiN \cr
& ~~~~~~~~~~~\times 
F_\al (\th_1 \cdots \th_N )_{\ep_1 \cdots \ep_N} 
F_{\al'}  (\th_N \cdots \th_1 )_{\ep_N^* \cdots \ep_1^*} 
\cr
&= \sum_{n=0}^\infty \inv{(n!)^2} 
\int d\th_1 \cdots d\th_{2n}  ~ 
\eui
\cr 
&~~~~~~~~~~~\times f_\al (\th_1 \cdots \th_{2n} ) 
f_{\al'}  (\th_{2n} \cdots \th_1 ) 
. \cr }}
In the second equation, we have set $N=2n$, and summed over
the $\left(\matrix{2n\cr n \cr}\right)$ possible 
permutations of $U(1)$ charges. 

We now show how these correlation functions may be
expressed as a Fredholm determinant:
\eqn\eIIxvii{
\lvac \sigma_\al (z , \zb ) \> \sigma_{\al'} (0) \rvac
= \Det ( 1 + {\bf K}^{(\al , \al' )} ) . }
Above, 
${\bf K}^{(\al , \al' )}$ is a $2\times 2$ matrix of 
Fredholm integral operators:
\eqn\ekk{
{\bf K} (u,v)  = \left(\matrix{K_{--} (u,v)  & K_{-+} (u,v) 
\cr K_{+-} (u,v)  & K_{++} (u,v)  \cr }\right)
. }  
For simplicity of notation,
we will not always display the $\al , \al'$ 
(and $z , \zb$ ) dependence of the kernel ${\bf K}$. 
The basic Fredholm theory of scalar integral operators
can be easily extended to finite matrices of 
integral operators.  One obtains 
\eqn\eIIxviii{
\Det (1 + {\bf K} ) 
= \sum_{N=0}^\infty \inv{N!} 
\sum_{\ep_1 \cdots \ep_N } 
\int_0^\infty \frac{du_1}{u_1} \cdots \frac{du_N}{u_N} 
\det \{ K_{\ep_i , \ep_j} (u_i , u_j ) \} , }
where 
$\det \{ K_{\ep_i , \ep_j } (u_i , u_j ) \} $ is 
an ordinary determinant of the finite $N\times N$ matrix
with $K_{\ep_i , \ep_j} (u_i , u_j ) $ as entries: 
\eqn\eIIxix{
\det \left\{ K_{\ep_i , \ep_j } (u_i , u_j ) \right\} 
=
\left\vert\matrix{ 
K_{\ep_1 \ep_1} (u_1 ,u_1 ) &K_{\ep_1 \ep_2} (u_1 ,u_2) &\cdots 
&K_{\ep_1 ,\ep_N} (u_1 , u_N ) \cr 
K_{\ep_2 \ep_1} (u_2 , u_1 ) &\cdot &\cdots &K_{\ep_2 \ep_N} (u_2 , u_N ) \cr
\cdot &\cdot &\cdots &\cdot \cr
\cdot &\cdot &\cdots &\cdot \cr
K_{\ep_N \ep_1} (u_N , u_1) &\cdot &\cdots &K_{\ep_N \ep_N} (u_N , u_N ) \cr} 
\right\vert . } 

With the appropriate choice of kernel ${\bf K}$, the expression
\eIIxviii\ coincides with the form factor sum \eIIxvi.  The
result is 
\eqn\eIIxx{\eqalign{ 
K_{++} (u,v) &= K_{--} (u,v) = 0 \cr 
K_{-+} (u,v) &= \frac{ e(u) \eh (v) }{u+v} \cr
K_{+-} (u,v) &= \frac{ \eh (u) e (v) }{u+v} \cr }}
where 
\eqn\eIIxxb{\eqalign{
e(u) &= \( \frac{\sin (\pi \al ) }{\pi} \)^{1/2} 
\> u^{(1+\al' - \al)/2} \> 
\eu \cr
\eh (u) &= \( \frac{\sin (\pi \al' ) }{\pi} \)^{1/2} 
\> u^{(1+\al - \al' )/2} \> 
\eu\cr
}}
The above choice of kernel is not the unique one that 
reproduces the expression \eIIxvi.  However, viewing 
\eIIxx\ as an ansatz, then the choice of $e , \eh$ is
unique up to constants. 

The result \eIIxx\eIIxxb\ is 
proven as follows. 
Note first that the condition $K_{++} = K_{--} = 0$ 
automatically projects the sum over $U(1)$ charges $\ep_i$ 
in \eIIxviii\ onto neutral states. 
Let 
\eqn\eIIxxi{
\{ K_{\ep_i \ep_j } (u_i , u_j) \} = 
\left(\matrix{ 0 &A \cr B & 0\cr }\right) , ~~~~~
\ep_i = -\ep_{i+n} = -1 , ~~i= 1, .. , n , } 
so that  
$$A = 
\left(\matrix{ 
K_{-+} (u_1 , v_1 ) &\cdots &K_{-+} (u_1 , v_n ) \cr 
K_{-+} (u_2 , v_1 ) &\cdots &K_{-+} (u_2 , v_n ) \cr 
\cdot &\cdots &\cdot\cr 
\cdot &\cdots &\cdot\cr 
K_{-+} (u_n , v_1 ) &\cdots &K_{-+} (u_n , v_n ) \cr }
\right) 
$$
\smallskip
$$
B = 
\left(\matrix{ 
K_{+-} (v_1 , u_1 ) &\cdots &K_{+-} (v_1 , u_n ) \cr 
K_{+-} (v_2 , u_1 ) &\cdots &K_{+-} (v_2 , u_n ) \cr 
\cdot &\cdots &\cdot\cr 
\cdot &\cdots &\cdot\cr 
K_{+-} (v_n , u_1 ) &\cdots &K_{+-} (v_n , u_n ) \cr }
\right) 
$$
where we have defined $v_i \equiv u_{i+n}$. 
Using the Cauchy identity,
\eqn\eIIxxv{
\det \( \inv{u_i + v_j} \) = \Delta (u_1 \cdots u_{2n} ) , }
we write the determinant of $\{K_{\ep_i\ep_j}(u_i,u_j)\}$ as
\eqn\eIIxxiv{ \eqalign{
\det \{K_{\ep_i\ep_j}(u_i,u_j)\} &=
\det \left(\matrix{0&A\cr B&0\cr }\right) 
= (-1)^n ~ \det A \det B \cr
&=(-1)^n \Delta (u_1 \cdots u_{2n} )^2\  
\prod_{i=1}^n \( e(u_i ) \eh (v_i ) \)^2 }}
Comparing \eIIxvi\ and \eIIxviii, and using the 
explicit expressions \eIIx~ of the form factors,
one sees that to prove eq. \eIIxvii~ we need to satisfy the relation 
\eqn\eIIxxvi{\eqalign{
(-1)^n \prod_{i=1}^n \( e(u_i ) \eh (v_i ) \)^2
&= \prod_{i=1}^n
\((u_i)^{\half-\al} (v_i)^{\half+\al}
 (u_i)^{\half+\al'} (v_i)^{\half-\al'} \) \cr
& ~~~~~~~~~~ \times  \( \exp (-mz(u_i +v_i ) -m\zb
 (u_i^{-1} + v_i^{-1} ) ) \) . \cr }}
The functions $e(u) , \eh (u)$  given in \eIIxxb~
were easily chosen to satisfy \eIIxxvi. 

Consider now a correlation function of the kind
\eqn\eIIs{
\bra{\Psi'}\sig_\al(z,\zb)\sig_{\al'}(0,0)\ket{\Psi},}
where $\ket{\Psi}$ and $\bra{\Psi'}$ are two $U(1)$-neutral
eigenstates of the Hamiltonian; i.e they are pure 
multi-particle states $\ket{\Psi}=
\ket{\beta_1,\cdots,\beta_N}_{\ep_1,\cdots,\ep_N}$ with
$\sum_i\ep_i=0$, and 
similarly for $\ket{\Psi'}$.
The correlation function  \eIIs~ can be computed using the crossing properties
and the factorized expression \eIIx~ of the form factors. One has~:
\eqn\eIIsa{\eqalign{
\bra{\Psi'}\sig_\al(z,\zb)\sig_{\al'}(0,0)\ket{\Psi}
= \sum_{n=0}^\infty & \inv{(n!)^2} 
\int d\th_1 \cdots d\th_{2n}  ~ \eui \cr 
& \times f_\al (\th_1 \cdots \th_{2n} ) 
f_{\al'}  (\th_{2n} \cdots \th_1 ) 
\prod_{i=1}^{2n} f_{\al;\al'}(\th_i;\Psi;\Psi')
, \cr }}
where $f_{\al;\al'}(\th;\Psi;\Psi')$ is a complicated function
depending on the rapidities $\beta$ and $\beta'$. 
Since the extra dependence on $\Psi$ and $\Psi'$ is factorized
into the product $\prod_i f_{\al;\al'}(\th_i;\Psi;\Psi')$,
it can be absorbed into a redefinition of the functions $e(u)$
and $\hat e(u)$. Thus one still has
\eqn\eIIsb{
\bra{\Psi'}\sig_\al(z,\zb)\sig_{\al'}(0,0)\ket{\Psi}
= {\rm Det}\(1 +{\bf K}_{\al,\al';\Psi,\Psi'}\),}
where ${\bf K}$ is still of the form \eIIxx, but with more
complicated functions $e,\ \hat e$. However, their $(z,\zb)$
dependence are identical. This is the origin of the fact 
that all these correlations satisfy the same differential equation.

\bigskip
\noindent
2.3 {\it Relation to ${\rm Ising}^2$ } 

\medskip

Since the Ising model is equivalent to a free real Majorana fermion,
whereas the $\sg$ theory is a free complex Dirac fermion, at the
lagrangian level, the $\sg$ theory is two copies of the Ising 
theory. It is well known that the above observation is somewhat
naive for a number of reasons.  For example, in the massless limit,
the Dirac and ${\rm Ising}^2$ theories differ in their detailed
partition functions on the torus\ref\rginsparg{P. Ginsparg, 
Les Houches Lectures 1988, in {\it Fields, Strings, and Critical
Phenomena}, E. Br\'ezin and J. Zinn-Justin Eds., North Holland (1990).}.  
For our purposes, the limitation of the identification of
$\sg$ with ${\rm Ising}^2$ is that only correlation functions
of certain special fields in the $\sg$ theory are related to 
${\rm Ising}^2$ correlation functions.  Let $\sigma (z , \zb )$
and $\mu (z , \zb )$ denote the Ising spin and disorder fields
respectively.  In \riz\rst\ the following
identifications were made:
\eqn\eIIxxxi{\eqalign{
\lvac \sigma (z ,\zb ) \sigma (0) \rvac^2  &= 
\lvac 
\sin (\Phi (z ,\zb )/2 ) \> \sin (\Phi (0)/2 ) \rvac
\cr 
\lvac \mu (z ,\zb ) \mu (0) \rvac^2  &= 
\lvac 
\cos (\Phi (z ,\zb )/2 ) \> \cos (\Phi (0)/2 ) \rvac, \cr}}
where on the RHS are $\sg$ correlation functions. 
In this section, we present a different derivation of part
of the relations \eIIxxxi\ based on  Fredholm determinants.

We first review some aspects of the form factor sum for the
Ising 
correlators\ref\rcm{J. L. Cardy and G. Mussardo, Nucl. Phys. B340
(1990) 387.}\ref\ryz{V. P. Yurov and Al. B. Zamolodchikov, Int. 
J. Mod. Phys. A6 (1991) 3419.}\rbb. 
Define 
\eqn\eIIxxxii{
\tau_\pm (z , \zb ) = \lvac \mu (z ,\zb ) \mu (0) \rvac 
\pm \lvac \sigma (z , \zb ) \sigma (0) \rvac . } 
In the free fermion description, the spectrum consists of a single
neutral particle.  From the known form factors of the spin and 
disorder fields, one obtains
\eqn\eIIxxxiv{
\tau_\pm (z , \zb ) 
= \sum_{n=0}^\infty \inv{n!} \( \frac{\pm 1}{2\pi} \)^n 
\int_0^\infty \frac{du_1}{u_1} \cdots \frac{du_n}{u_n} ~ 
\prod_{i<j} \( \frac{u_i - u_j}{u_i + u_j} \)^2 
 \eui .} 
The functions $\tau_\pm$ may be expressed as a Fredholm determinant
for scalar integral operators: 
\eqn\eIIxxxivb{
\tau_\pm = \Det ( 1\pm V) , } 
where $V$ is the kernel  
\eqn\eIIxxxivc{
V(u,v) = \inv{2\pi} \( \frac{2 \sqrt{uv}}{u+v} \) ~
\eu . } 

In terms of the functions $\tau_\pm$, the relations 
\eIIxxxi\ read 
\eqn\eIIxxxiii{\eqalign{ 
\lvac \sigma_{1/2} (z , \zb ) \sigma_{1/2} (0) \rvac 
&= \tau_+ \tau_- 
\cr 
\lvac \sigma_{1/2} (z , \zb ) \sigma_{-1/2} (0) \rvac 
&= \inv{2} \( \tau_+^2 + \tau_-^2 \) , \cr }}
where we have used the hermiticity relations 
$\langle \sigma_{\al} \sigma_{\al'} \rangle = \langle 
\sigma_{-\al} \sigma_{-\al'} \rangle$. 
There is a simple proof of the first relation in 
\eIIxxxiii\ based on  Fredholm determinants. 
For the particular correlator $\langle \sigma_{1/2} \sigma_{1/2} 
\rangle $, 
from \eIIxx\eIIxxb\ one finds that 
$K_{+-} = K_{-+} = V$, so that 
\eqn\eIIxxxvi{
\Det \( 1 + {\bf K^{(1/2,1/2)} } \) = 
\Det \left(\matrix{1 &V\cr V &1\cr}\right) 
= \Det (1+V) \> \Det (1-V) = \tau_+ \tau_-  
} 
The second relation in \eIIxxxiii\ is more non-trivial, 
since it relates  a Fredholm determinant to a sum
of such determinants.  We have not found a simple proof of
it, however we have verified it to lowest non-trivial order. 

In the next section, we will derive differential equations
satisfied by the logarithm of the $\sg$ correlation functions.
For $\langle  \sigma_{1/2} \sigma_{1/2} \rangle$, these differential 
equations can obviously be derived from the known Ising differential
equation, since $\log (\tau_+ \tau_- ) = \log \tau_+ + \log \tau_-$. 
However, for $\langle \sigma_{1/2} \sigma_{-1/2} \rangle$, it is not evident
how to obtain our result from the known Ising results. 

The methods we develop in the next sections can be used
to provide a new proof of the known differential equations
which govern the Ising correlation functions;  this is
described in appendix D. 

\def\dpsi{ {\bf \Psi}}
\def\zb{ {\bar z}}
\def\cpm{C_\pm(\th)}
\def\apm{C_{\pm}^{\dag}(\th)}

\newsec{ Differential Equations for Correlation Functions.}

In this section we prove the result stated in the introduction
by using the Fredholm determinant expressions for the correlation
functions derived in the last section. The method we follow is
a generalization of methods developed in \rkor\book. 
As will be clear in the sequel, the proof does not rely on
the detailed properties of the kernel ${\bf K}$, such as its
$\al$ dependence, but rather only on its most salient features.  

The $z,\ \zb$ dependence of all the quantities we will introduce
are a consequence of 
the $z,\ \zb$ dependence of the functions $e(u)$ and $\hat e(u)$.
The latter are very simple: 
\eqn\EDk{\eqalign{
 \d_z e(u) &= -\frac{mu}{2} e(u) , \quad
\d_z \hat e(u) = -\frac{mu}{2} \hat e(u) \cr
 \d_\zb e(u) &= -\frac{m}{2} e '(u) , \quad
\d_\zb \hat e(u) = -\frac{m}{2} \ehp (u)  , \cr}} 
where for convenience we have defined:
\eqn\eepr{
e' (u) \equiv \frac{e (u)}{u} , \quad 
\ehp (u) \equiv \frac{\eh (u)}{u} . } 
Expressions for derivatives of $\kb$  follow from eqs. \EDk: 
\eqn\EDjj{\eqalign{
\d_z K_{-+} (u,v) &= -\frac{m}{2} e(u)\ \hat e(v),\quad
\d_z K_{+-} (u,v) = -\frac{m}{2} \hat e(u)\ e(v) \cr
\d_\zb K_{-+} (u,v) &= -\frac{m}{2} e'(u)\ \ehp (v),\quad
\d_\zb K_{+-} (u,v)  = -\frac{m}{2} \ehp (u)\ e'(v) . \cr
}}
Thus, one sees  that $\d_z \kb, ~ \d_\zb \kb $ are projectors. 

\medskip
\noindent 
3.1 {\it The resolvent} 

\def\kb{{\bf K}}
\def\rb{{\bf R}}

We will need the resolvent ${\bf R}$ for the kernel ${\bf K}$,
defined by the matrix equation   
\eqn\EDa{
(1-{\bf R}) (1 + \kb )  =1 . }
As for ${\bf K}$, the resolvent ${\bf R}$ is  a 
$2\times 2$ matrix of integral operators. 

For simplicity, consider first scalar kernels of the 
following form:
\eqn\EBa{
V_{(\pm )} (u,v) = \frac{ \sum_a e_a(u)e^a(v)}{u\pm v} }
The index $(\pm )$ refers to the nature of the simple
pole of $V_{(\pm )}$.  Kernels of this type form a group
with a $\Zmath_2$ graded multiplication law.  Let
$[ V_{(\pm)} ]$ denote the class of kernels of the type 
$V_{(\pm)}$.  Then using the identity 
\eqn\eIIIi{
\inv{u+w} \inv{w+v} = \inv{u-v} 
\( \inv{w+v} - \inv{u+w} \) , } 
one has the following multiplication law: 
\eqn\eIIIii{
[V_{(-)} ] \times [V_{(-)} ] = [V_{(-)} ] , ~~~~~
[V_{(+)} ] \times [V_{(-)} ] = [V_{(+)} ] , ~~~~~
[V_{(+)} ] \times [V_{(+)} ] = [V_{(-)} ] . } 
Scalar kernels are multiplied in the conventional manner:
\eqn\emult{
\( V^{(1)} V^{(2)} \) (u,v) = \int_0^\infty \frac{dw}{w} 
V^{(1)} (u,w) V^{(2)} (w,v) . }
For example 
\eqn\EBd{
\({ V_{(+)}^{(1)}V_{(+)}^{(2)}}\)(u,v) =
\frac{\sum_a e_a^{(1)}(u)\({e^{a,(1)}V_{(+)}^{(2)}}\)(v)
- \({V_{(+)}^{(1)}e_a^{(2)}}\)(u)e^{a,(2)}(v)}{u- v}  . }
This generalizes the study of \rkor\book, where only kernels of
type $V_{(-)}$ were considered, which by themselves form
a group.  
This group property is important 
since it ensures that the resolvent is also an element of the
group. In Appendix D, further properties of these kernels 
and their resolvents are presented. 

Let us return now to the construction of $\rb$.  For future
convenience, consider the slightly more general problem of
finding the resolvent ${\bf R_\pm}$ for the kernel $\pm \kb$: 
\eqn\eIIIiii{
(1 - {\bf R_\pm} ) (1 \pm \kb ) = 1, }
where ${\bf R} = {\bf R_+}$.  Expressing 
\eqn\EDb{
{\bf R_\pm}  = \pmatrix{ F & \pm D \cr \pm E & G \cr} }
where $D,\ E,\ F,\ G$ are scalar kernels, 
equation \eIIIiii\ is equivalent to the system of equations
\eqn\EDc{\eqalign{
F+DK_{+-} = 0 &,\quad K_{-+}=D + F K_{-+} \cr
G + EK_{-+} =0 &,\quad K_{+-}= E+GK_{+-}  . \cr}}
Due to the fact that $\kb = \kb^t$, one has
$F^t = F , ~ D^t = E , ~ G^t = G$, where $t$ denotes 
transpose of matrices and kernels.  
Since $K_{+-}$ and $K_{-+}$ are of type $[V_{(+)} ]$, one deduces
that $F,G \in [V_{(-)} ]$ and $D,E \in [V_{(+)} ]$.  

\def\xp{ \xi^+}  \def\xm{ \xi^- }
\def\xph{\hat{\xi}^+}  \def\xmh{\hat{\xi}^- }
\def\xpm{\xi^\pm} 
\def\xpmh{\hat{\xi}^\pm} 
\def\eh{{\hat{e}}}

The equations
\EDc\ can be used to express $\rb$ in terms of $\kb$.  We first
state the result then describe how to prove it. 
Define the functions $\xpm , ~ \xpmh$, where
\eqn\eIIIiv{
\col = 
2 (1 - {\bf R_\pm} ) ~ \left(\matrix{e \cr \eh \cr}\right) , }
so that 
\eqn\EDe{\eqalign{
\xp = 2(1-F)e &,\quad \xm = - 2D \hat e \cr
\xph =2 (1-G)\hat e &,\quad \xmh = -2 E e .\cr}}
The $\xpm , \xpmh$ are functions of $z , \zb$, and $u$. 
Functions  are multiplied by integral operators  
in the same way as in \emult, e.g. 
\eqn\vmult{
(Fe) (u) = \int_0^\infty \frac{dv}{v} F(u,v) e(v)  . } 
Then $\rb$ can be expressed in terms of these functions: 
\eqn\EDd{\eqalign{
D(u,v)&= \frac{\xp(u)\xph(v)-\xm(u)\xmh(v)}{4(u+v)}\cr
E(u,v)&= \frac{\xph(u)\xp(v)-\xmh(u)\xm(v)}{4(u+v)}\cr
F(u,v)&= \frac{\xp(u)\xm(v)-\xm(u)\xp(v)}{4(u-v)}\cr
G(u,v)&= \frac{\xph(u)\xmh(v)-\xmh(u)\xph(v)}{4(u-v)} .\cr}}

For future reference we will also need 
\eqn\eIIIv{\eqalign{
2 e &= \xp + \xmh K_{+-} ,~~~~~ 0 = \xm + \xph K_{+-} \cr
2 \eh &= \xph + \xm K_{-+} ,~~~~~ 0 = \xmh + \xp K_{-+} .\cr}}
These relations are the component form of the matrix relation
\eqn\eIIIvi{
\left(\matrix{e\cr \eh \cr }\right)^t 
= \( \inv{2} (1 \pm \kb ) 
\col  \)^t ,}
the latter being a simple consequence of \eIIIiv. 

The proof that the resolvent takes the form indicated in
\EDd\ is by direct computation.  We illustrate this
by proving the first relation in \EDc, $F+D K_{+-} = 0$,
the other relations being proved similarly. Using the
expression for $D$ in \EDd, one obtains 
\eqn\eIIIvii{
( D K_{+-} ) (u,v) 
= \inv{4} \int \frac{dw}{w} 
\( \inv{u+w}  \) 
\( \xp (u) \xph (w) - \xm (u) \xmh (w) \) \( \inv{w+v} \) 
\eh (w) e(v).} 
Using now the identity \eIIIi, this becomes 
\eqn\eIIIviii{
( D K_{+-} ) (u,v) =
\inv{u-v} 
\( - (D \eh ) (u) e(v) + \inv{4} \xp (u) ( \xph K_{+-} )(v) 
 - \inv{4} \xm (u) ( \xmh K_{+-} ) (v)  \). }
 The relations \EDe\ and \eIIIv\ can now be used to show
 that the RHS of \eIIIviii\ equals $-F$.

\medskip\noindent
3.2 {\it The potentials}

We now have to define certain functions  which
are essential for the proof, since they are directly
related to the derivatives of the logarithm of the
correlation functions. Define the ``potentials" 
\eqn\EDf{\eqalign{
B_\pm = \vev{e,\xi^\pm}-(1 \mp 1)&,\quad 
\hat B_\pm=\vev{\hat e,\hat \xi^\pm}-(1 \mp 1)\cr
C_\pm = \vev{e',\xi^\pm}-(1 \mp 1)
&,\quad 
\hat C_\pm=\vev{\ehp ,\hat \xi^\pm}-(1 \mp 1)\cr}}
where $\vev{~,~}$ denotes the scalar product defined for any functions
$a(u)$ and $b(u)$ by 
\eqn\EDg{
\vev{a,b} = \int_0^\infty \frac{du}{u} a(u)b(u) .}
The potentials are functions only of $z , \zb$. 
They are not all independent due to the symmetry properties of ${\bf K}$.
We have the following algebraic relations~:
\eqn\EDh{\eqalign{
& B_- = \hat B_- \cr
& C_-\hat C_- - C_+\hat C_+ = 4 \cr}}
The first relation is easily proved using $D^t=E$:
\eqn\econs{
B_- = -2 \vev{e, D\eh } = -2 \vev{e,D \eh}^t = -2 \vev{\eh , Ee} 
= \hat{B}_- ~. } 
The second relation is proven in appendix B. 

Let $\Sig(z,\bar z)$ denote the logarithm of the Fredholm
determinant~:
\eqn\EDi{
\Sig(z,\bar z) = \log {\rm Det}(1+ {\bf K}) }
Using $\log {\rm Det}(1+ {\bf K})= {\rm Tr}\log (1+ {\bf K})$,
its derivative with respect to $z$ is expressed in terms of the resolvent~:
\eqn\EDj{
\d_z \Sig(z,\bar z) = {\rm Tr}\[{(1-{\bf R}) \d_z {\bf K}}\]}
The trace can easily be computed since $(\d_z {\bf K})$
is a projector. From \EDjj, \EDe\ one obtains:  
\eqn\EDl{
\d_z \Sig(z,\bar z)= - \frac{m}{4} \({ B_- + \hat B_- + 4 }\) = 
- \frac{m}{2}(B_- + 2) }

To find a differential equation obeyed by $\Sig$, one next has
to derive differential equations for the potentials.  This
will be done in the next sections. 

\vfill\eject

\noindent
3.3 {\it The auxiliary linear system} 

Non-linear differential
equations for the potentials can be derived 
by introducing an auxiliary linear system
of differential equations. 
This system is obtained by computing  
 the derivatives of the two dimensional vectors ${\bf \xi}$ and
${\bf {\hat \xi}}$ with components $\xi^\pm$ and $\hat \xi^\pm$~:
\eqn\EDq{
{\bf \xi}(u)=\pmatrix{ \xm(u) \cr \xp(u) },\quad
{\bf \hat \xi}(u)=\pmatrix{ \xmh(u) \cr \xph(u) } . }
The $z, \ \zb$  derivatives of these vector functions 
are computed using their explicit definition
in terms of the components $D,\ E,\ F,\ G$ of the resolvent \EDd~
and the formula \EDk~ for the derivatives of the functions
$e(u)$ and $\hat e(u)$. One finds that~:
\eqn\EDr{
\cases{\d_z\ {\bf \xi} = A_z {\bf \xi} &\cr ~&~\cr
\d_\zb\ {\bf \xi} = A_\zb {\bf \xi} &\cr },\quad
\cases{\d_z\ {\bf \hat \xi} = \hat A_z {\bf \hat \xi} &\cr ~&~\cr
\d_\zb\ {\bf \hat \xi} = \hat A_\zb {\bf \hat \xi} &\cr } }
with
\eqn\EDs{\eqalign{
A_z &= \frac{m}{2} \pmatrix{ u & \hat B_+ \cr
			B_+ & -u \cr} \cr ~&~\cr
A_\zb & = \frac{m}{8u}
\pmatrix{ C_-\hat C_- + C_+ \hat C_+ & -2 C_- \hat C_+ \cr
	 2 C_+ \hat C_- & - C_-\hat C_- - C_+ \hat C_+ \cr} \cr}}
The matrices $\hat A_z$ and $\hat A_\zb$ have the same expressions
as $A_z$ and $A_\zb$ but with ``hatted" and ``unhatted" potentials
interchanged.
The proof of these equations is given in Appendix C.

\medskip\noindent
3.4 {\it The differential equations} 

 From the linear system \EDr, one can deduce differential equations
satisfied by the potentials.  The result is the following: 
\eqn\EDu{\eqalign{
\d_z C_+ &= \frac{m}{2} B_+ C_-,\quad 
\d_z C_- = \frac{m}{2} \hat B_+ C_+\cr 
\d_z \hat C_+ &= \frac{m}{2} \hat B_+ \hat C_-,\quad 
\d_z \hat C_- = \frac{m}{2}  B_+ \hat C_+\cr 
\d_\zb B_+ &= \frac{m}{2} C_+ \hat C_- ,\quad
\d_\zb \hat B_+= \frac{m}{2} \hat C_+ C_- \cr
\d_\zb B_- &= \frac{m}{4}  \( \hat C_+ C_+ + \hat C_- C_- - 4 \) 
= \frac{m}{2} \hat C_+ C_+ . \cr 
}}

Since all of the equations in \EDu\ are proven similarly,
we illustrate the proof by example. 
 From the definition of $B_+$, one has 
\eqn\addi{
\d_\zb B_+ = \vev{\d_\zb e , \xp } + \vev{ e , \d_\zb \xp } . } 
Substituting \EDk\ and $\d_\zb \xp$ from \EDr, one finds 
\eqn\addii{
\d_\zb B_+ = - \frac{m}{2} \vev{e' , \xp} - \frac{m}{8} 
(C_+ \Ch_+ + C_- \Ch_- ) \vev{e, \frac{\xp}{u} } 
+ \frac{m}{4} C_+ \Ch_- \vev{e, \frac{\xm}{u} } . } 
Noting that $\vev{e, \xpm /u } = \vev{e' , \xpm } $, and using 
the definitions \EDf, one obtains the result in \EDu. 
Consider next $\d_z C_+$:
\eqn\addiii{\eqalign{ 
\d_z C_+ &= \vev{ \d_z e' , \xp} + \vev{e' , \d_z \xp } \cr
&= - \frac{m}{2} \vev{e , \xp} + \frac{m}{2} B_+ 
\vev{e' , \xm } - \frac{m}{2} \vev{e' , u \xp } . \cr }}
Noting 
$\vev{e' , u \xp } = \vev{ e , \xp} $, one obtains the
desired result. 

One can also find  
differential equations for the potentials from the zero-curvature
condition 
for the 
connection $A_z,\ A_\zb$ ~:
\eqn\EDt{
\[{\ \d_z+A_z\ ,\ \d_\zb + A_\zb\ }\] =0,  }
and similary for the connection $\hat A_z,\ \hat A_\zb$. 
These zero curvature conditions yield a set of differential
equations which is in fact weaker than, but compatible with, 
the set of equations \EDu.  Since we will not need the
equations which follow from the zero-curvature condition, we
do not present the details. 

The last equation in \EDu, along with \EDl,
can be used to express the derivatives
of the logarithm of the correlation functions only in terms of the
$C$ potentials:
\eqn\EDext{
\d_\zb \d_z\ \Sig(z,\zb) = -\frac{m^2}{4} \hat C_+ C_+ . }

Notice that the $B$ potentials can be eliminated from eqs. \EDu.
This leads to differential equations involving only the $C$
potentials~:
\eqn\EDv{\eqalign{
&C_-\d_z \hat C_- = \hat C_+ \d_z C_+ \quad ,\quad
C_+\d_z \hat C_+ = \hat C_- \d_z C_- \cr
&C_+(\d_z\d_\zb C_-) - (\d_z C_-)(\d_\zb C_+) = 
\frac{m^2}{4}\ C_-C_+ (C_+\hat C_+)  \cr
&C_-(\d_z\d_\zb C_+) - (\d_z C_+)(\d_\zb C_-) = 
\frac{m^2}{4}\ C_-C_+ (C_-\hat C_-)  \cr}}
Two additional equations corresponding to interchanging 
$C_\pm $ with $\hat C_\pm $ are derived similarly. 
These equations are similar
but not identical to the Hirota equation for the $\hat{sl_2}$ affine hierarchy.

In order 
to solve \EDu\ and \EDv,  we introduce a parameterization 
of the $C_\pm, \hat C_\pm$ potentials
which solves the algebraic constraint \EDh~:
\eqn\EDw{\eqalign{
C_- = 2 e^a\ \cosh\varphi &,\quad C_+ = 2e^b\ \sinh\varphi \cr
\hat C_- = 2 e^{-a}\ \cosh\varphi &,\quad \hat C_+ = 2e^{-b}\ \sinh\varphi \cr}}
with $a,\ b$ and $\varphi$ auxiliary functions, depending on $z, \ \zb$.
Inserting  this parameterization \EDw~ into the differential equations
gives the following.  The first two equations in \EDv\ give
\eqn\emodi{
\d_z a = -\tanh^2 \varphi ~ \d_z b . }
Using this equation and its $\d_\zb$ derivative the second 2 equations
can be simplified to 
\eqn\emodii{\eqalign{
( \d_z \d_\zb a )  ~ \coth \varphi  - ( \d_z \d_\zb b ) ~\tanh \varphi 
-2 \d_z \varphi ~ \d_\zb (b-a ) =0 \cr
\d_z \d_\zb \varphi = \frac{m^2}{2} \sinh 2\varphi 
- \tanh \varphi ~ \d_z b ~ \d_\zb (b-a) . \cr }}

The function $b$ can be deduced using Lorentz invariance.   
Let $z = r e^{i \theta} /2 $, $\zb = r e^{-i \theta} /2 $, and
consider shifts of $\theta$ by $\gamma$. 
The functions $e, \hat{e}$ satisfy
\eqn\emodiii{\eqalign{
e( e^{i\gamma} z , e^{-i \gamma} \zb , u ) 
&= e^{-i\gamma ( 1 + \alpha' - \alpha )/2 } e( z, \zb , e^{i\gamma} u ) \cr
\hat{e} ( e^{i\gamma} z , e^{-i \gamma} \zb , u ) 
&= e^{-i\gamma ( 1 + \alpha - \alpha' )/2 } e( z, \zb , e^{i\gamma} u ) 
. \cr }}
From the definition \EDf\ of $C_+, \hat{C_+}$, and making the change
of variables $u \to e^{-i\gamma} u$, one finds 
\eqn\emodmod{
C_+ = e^{2i(\alpha - \alpha') \theta}  f(r) ,
~~~~~
\hat{C}_+ = e^{-2i(\alpha - \alpha') \theta}  \hat{f}(r) ,
}
for some scalar functions $f , \hat{f}$.   Then, using
\eqn\emodxx{
e(z, \zb , u) = u ~ \hat{e} (\zb, z, 1/u ) , }
one can show $f = \hat{f}$ by making the change of variables $u\to 1/u$.  
Thus, 
\eqn\emodiv{
e^{2b} = \frac{C_+}{\hat{C}_+} = 
e^{4 i (\alpha - \alpha' ) \theta}
, } and 
\eqn\emodv{  
b = (\alpha - \alpha') \log \( \frac{z}{\zb} \) . }
Inserting this $b$ into \emodi\ and taking the complex conjugate, one
deduces
\eqn\emodvi{
\d_\zb a = \tanh^2 \varphi ~ \d_\zb b . } 
The function $\varphi$ is only a function of $r$.  Thus
\emodii\ can be written as
\eqn\resulti{
\( \d_r^2 + \inv{r} \d_r \)  \varphi = \frac{m^2}{2} \sinh 2\varphi 
+ \frac{4 (\alpha - \alpha')^2 }{r^2} \tanh \varphi (1-\tanh^2 \varphi )
. }

Finally, using the equation \EDext, and also \EDw, one obtains 
\eqn\EDx{
\( \d_r^2 + \inv{r} \d_r \)  \Sig(r) = - m^2\ \sinh^2\varphi 
=  \frac{m^2}{2} \({1 - \cosh 2\varphi   }\) . }
This is the result announced in the introduction. 
Notice that $\d_z \d_\zb \Sig$ is only parameterized by a single
function $\varphi (r)$, and the differential equation for
$\varphi$ involves only $\varphi$ itself. 

\newsec{Concluding Remarks}

Though the proof is somewhat lengthy, the main result stated
in the introduction is strikingly simple.  
What about away from the free fermion point? 
Certainly, our derivation relies essentially on the fact
that the particles in the form factor sum are free.  
This fact is ultimately responsible for the properties
of the form factors that allowed us to express the form
factor sum as a Fredholm determinant. 
Nevertheless, the simplicity of the final result suggests
it is worthwhile to investigate whether other derivations of it
are possible that make no reference to the free fermions.   
For instance, the $SG_{ff}$ theory possesses an 
exact $\hat{sl(2)}$ affine symmetry\rlec, which is known
to become $q$ deformed away from the free fermion 
point\ref\rbl{D. Bernard and A. LeClair, Commun. Math. Phys. 142
(1991) 99.}. 
It would be very interesting to be able to derive the result
presented in this paper solely from the $\hat{sl(2)}$ 
symmetry, as such a derivation may be amenable to $q$-deformation.

\vfill\eject

\centerline{\bf Acknowledgements}

We would like to thank O. Babelon, 
R. Konik,  V. Korepin, and N. Slavnov for discussions. 
We are also grateful to the Aspen Center for Physics for the opportunity
to begin this work. A.L. is supported by an Alfred. P. Sloan Foundation
Fellowship, and the National Science Foundation in part through
the National Young Investigator program. 

\bigskip

\appendix{A}{Form factors from monodromy.}

Here we describe how the form factors in section 2  
can be derived from  braiding relations with the fermions. 
As in  the ising case\rmjs, 
the braiding relations induce a Bogoliubov
transformation on the creation-annihilation operators
in  momentum space. The field, which is the unitary operator implementing
this transformation, is therefore the exponential of
a bilinear form in the fermions.

Let  $\dpsi_\pm=\pmatrix{ {\bar \psi}_\pm \cr \psi_\pm \cr}$
be the Dirac field satisfying the Dirac equation:
\eqn\Aa{
\d_z {\bar \psi}_\pm = im \psi_\pm , \quad
\d_{\zb} \psi_\pm = -im {\bar \psi}_\pm }
Its mode decomposition in momentum space is:
\eqn\Ab{\eqalign{
\psi_\pm(x,t)&=\pm \sqrt{m} \int d\th e^{\th/2} 
\({ C_\pm(\th) e^{-ip\cdot x} - C_{\mp}^{\dag} (\th) e^{ip\cdot x} }\) \cr
\bar \psi_\pm(x,t)&=\mp i \sqrt{m} \int d\th e^{-\th/2} 
\({ C_\pm(\th) e^{-ip\cdot x} + C_{\mp}^{\dag} (\th) e^{ip\cdot x} }\) \cr}}
with $p\cdot x = m(x\sinh\th  -t  \cosh\th)$.

Let $\sig_\al = : \exp\({i\al \Phi(0,0) }\) :$. 
This field  is non-local with respect to the fermions, and therefore
has non-trivial equal time braiding relations with them:
\eqn\Ac{\eqalign{
& \sig_\al\ \dpsi_\pm(y) \ \sig_\al^{-1} = \dpsi_\pm(y),
\quad \quad ~~~~~~~\quad {\rm for}\ y<0 \cr
& \sig_\al\ \dpsi_\pm(y) \ \sig_\al^{-1} = 
e^{\pm i 2\pi\al}\ \dpsi_\pm(y),
\quad \quad {\rm for}\ y>0 \cr}}
Here, the fermions $\dpsi_\pm(y)$ are at $t=0$.

Equation \Ac~ induces a Bogoliubov transformation on the $\cpm$'s
and $\apm$'s. To describe this transformation, one has to Fourier transform 
$\sig_\al\dpsi_\pm(y)\sig_\al^{-1}$. 
Decomposing it into the sum of two integrals, one from $-\infty$ to zero,
the other from zero to $+\infty$, and using the braiding relation \Ac,
we get:
\eqn\Ad{\eqalign{
\int_{-\infty}^{+\infty} \frac{dy}{2\pi} e^{iky} 
&\({ \sig_\al\ \dpsi_\pm(y) \ \sig_\al^{-1} }\) \cr
&= \int_{-\infty}^{+\infty} \frac{dy}{2\pi} e^{iky} \dpsi_\pm(y)
+ \({ e^{\pm i2\pi \al}-1 }\) 
\int_0^{+\infty} \frac{dy}{2\pi} e^{iky} \dpsi_\pm(y) \cr}} 
Paramatrizing $k$ as $k=m\sinh\beta$, the integrals are easily done.
We find:
\eqn\Ae{
\({\sig_\al\ C_\pm(\beta) \ \sig_\al^{-1}}\) = C_\pm(\beta)
+ \({\frac{1-e^{\pm i2\pi\al}}{4i\pi}}\) \int_{-\infty}^{+\infty}
d\th \[{ \frac{C_\pm(\th)}{\sinh(\frac{\beta-\th+i\ep}{2})}
- \frac{C_{\mp}^{\dag}(\th)}{\cosh(\frac{\beta-\th}{2})} }\] }
Here $\ep \to 0^+$. A similar relation holds for
$C_{\pm}^{\dag}(\beta)$ since $\sig_\al$ is a unitary
operator.

Equation \Ae~ is a linear transformation of the creation-annihilation
operators which preserves their commutation relations, since it
is defined by conjugation. Therefore it is a Bogoliubov
transformation. We need a few basic facts about these
transformations, cf. \ref\bogo{F. A. Berezin, {\it Method
of Secondary Quantization}, Nauka, Moscow, 1965.}. 
Let $C_i$ and $C^{\dag}_i$ be fermionic operators with
$\{C_i,C^{\dag}_j\}=\delta_{ij}$. Consider the 
linear transformation $C_i\to D_i$ specified by:
\eqn\Af{\eqalign{
D_i &= \sum_j\({ C_j {\bf X}_{ji} + C^{\dag}_j {\bf Y}_{ji} }\) \cr
D^{\dag}_i &= \sum_j\({ C^{\dag}_j {\bar {\bf X}}_{ji} + C_j {\bar {\bf Y}}_{ji} }\) \cr}}
This  is a canonical transformation if the operators
$D_i$ and $D^{\dag}_i$ also satisfy the fermionic relations
$\{D_i,D^{\dag}_j\}=\delta_{ij}$. This amounts to a few constraints
on the matrices ${\bf X}$ and ${\bf Y}$. The unitary operator
$U$ which implements this transformation, i.e.
$D_i = U  C_i  U^{-1}$, is given by:
\eqn\Ah{
U= \inv{\CN} :\exp\({\half\sum_{ij}\({ C^{\dag}_j {\bf F^c}_{ji} C^{\dag}_i
+ 2 C^{\dag}_j {\bf R}_{ji} C_i + C_j {\bf F}_{ji} C_i }\)}\) :}
where $\CN=\[{\rm Det}({\bf X}^{\dag}{\bf X})\]^{-1/4}$. 
The unitary operator \Ah~ exists whenever this normalisation constant
is finite. In eq. \Ah, the matrices ${\bf F^c}$, ${\bf R}$ and
${\bf F}$ are defined by:
\eqn\Ai{
{\bf F^c}={\bf Y}(1 + {\bf R}),\quad
{\bf X}(1+ {\bf R}) = 1,\quad
{\bf F} = (1+{\bf R}){\bar {\bf Y}} }
These matrices are the two-particle form factors of the operator $U$,
e.g.
\eqn\Ax{
\vev{U C^{\dag}_jC^{\dag}_i} = \inv{\CN}\ {\bf F}_{ij} }
Therefore, the operator $U$ is completely characterized by its
two-particle form factors.

In our case, the fermion operators $C_\nu(\th)$ have a discrete index
$\nu=\pm$ and a continuous index $\th$ taking values from $-\infty$
to $+\infty$. The matrices ${\bf X}$ and ${\bf Y}$ are therefore
kernels. From eq. \Ae, we deduce that:
\eqn\Ak{\eqalign{
{\bf X}^{\nu,\nu'}(\th,\th') &= \delta^{\nu,\nu'}\({
\delta(\th-\th') + V_\nu(\th,\th') }\) \cr
{\bf Y}^{\nu,\nu'}(\th,\th') &= \delta^{\nu,-\nu'}
Y_\nu(\th,\th') \cr}}
with
\eqn\Al{\eqalign{
V_\nu(\th,\th')&= \({\frac{e^{i\nu 2\pi\al}-1}{4\pi i}}\)
\inv{ \sinh(\frac{\th-\th'-i\ep}{2} )} \cr
Y_\nu(\th,\th')&= \({\frac{e^{-i\nu 2\pi\al}-1}{4\pi i}}\)
\inv{ \cosh(\frac{\th-\th'}{2})} \cr}}

To find the expression of the ``spin" field $\sig_\al$,
we first have to find the resolvent ${\bf R}$. It is of the form
\eqn\Axx{
{\bf R}^{\nu,\nu'}(\th,\th')
= \delta^{\nu,\nu'}\ R_\nu(\th,\th')}
The kernel $R_\nu$ is the resolvent of $V_\nu$:
\eqn\An{ (1+V_\nu)(1+R_\nu)= 1}
We claim that a solution to \An~ is ~:
\eqn\Am{
R_\nu(\th,\th')= \({\frac{1-e^{i\nu 2\pi\al}}{4\pi i}}\)
\frac{e^{-\nu(\th-\th')}}{\sinh(\frac{\th-\th'+i\ep}{2})} }

To prove this, we have to compute the product
$(V_\nu R_\nu)$:
\eqn\Amm{
(V_\nu R_\nu ) (\th , \th ' ) 
= \int_{-\infty}^\infty d\beta 
 \> V_\nu (\theta, \beta ) R_\nu (\beta , \th' ). }
This involves an integral over
the intermediate rapidity $\beta$ from $-\infty$ to
$+\infty$. Since the function to integrate is  $(i2\pi)$-periodic
in $\beta$ up to a phase, which is equal to $e^{-i\nu2\pi\al}$,
we can compute its integral by considering
the contour in the $\beta$ complex plane  to be the
sum of 2 contours, one along the real axis at ${\rm Im} (\beta)
= 0$, the other along the line ${\rm Im} (\beta) = -2\pi i$.
The integral is then given by the residue theorem; the poles,
inside the contour, which are all simple poles,
are at $\beta=\th-i\ep$ and at $\beta=\th'-i\ep$.
Suming the residue contributions gives:
\eqn\Ap{
V_\nu\ R_\nu + V_\nu +R_\nu = 0 }
as claimed.

The matrix kernel ${\bf F}$ has the following form:
\eqn\Azz{ 
{\bf F}^{\nu,\nu'}(\th,\th')=\delta^{-\nu,\nu'}\ F_\nu(\th,\th')}
with $F_\nu = (1+R_\nu) {\bar Y}_\nu$. They are computed in a 
similar way using contour integrals. We find
\eqn\Aq{
F_\nu(\th,\th')=\nu\ \frac{\sin(\pi\al)}{2i\pi}\
\frac{e^{-\nu(\th-\th')}}{\cosh(\frac{\th-\th'}{2})} }

This coincides with the two particle
form factors given in the text. The crossed form
factors $F^c$ are derived in the same way.

\def\zp{\zeta^+}
\def\zm{\zeta^-}
\def\zpm{\zeta^\pm}
\def\zph{\hat{\zeta}^+}
\def\zmh{\hat{\zeta}^-} 
\def\zpmh{\hat{\zeta}^\pm} 
\def\Ch{\hat{C}}

\appendix{B}{Relations between the potentials.}

Let the functions $e' , \ehp $ of $z ,\zb $ and $u$ 
be as defined in \EDk.  Analagously to \EDe, define 
additional vector functions $\zpm ,\ \zpmh $ 
of $z, \zb$ and $u$ as follows: 
\eqn\eBi{\eqalign{
\zp = 2(1-F)e'  &,\quad \zm = - 2D \ehp \cr
\zph =2 (1-G)\ehp &,\quad \zmh = -2 E e' .\cr}}
These functions can be linearly related to the $\xpm ,\ \xpmh$
functions, the coefficients being the potentials $C_\pm 
, \hat{C}_\pm$.  The result is 
\eqn\eBii{\eqalign{
\zm (u) &= \inv{2u} \( 
\Ch_- \xm (u) - \Ch_+ \xp (u) \) \cr  
\zp (u) &= \inv{2u} \( 
-C_- \xp (u) + C_+   \xm (u) \) \cr  
\zmh (u) &= \inv{2u} \( 
C_- \xmh (u) - C_+ \xph (u) \) \cr  
\zph (u) &= \inv{2u} \( 
-\Ch_- \xph (u) + \Ch_+ \xmh (u) \)  .\cr}}

All of the above relations are proven similarly, so we illustrate
the technique with one of them.  From the definition \eBi, and the
explicit form of $D$ in \EDd, one has 
\eqn\eBiii{
\zm (u) = - \inv{2} \int \frac{dw}{w} 
\( \inv{u+w} \inv{w} \) 
\( \xp (u) \xph (w) - \xm (u) \xmh (w) \) \eh (w) . } 
Using the relation \eIIIi\ with $v=0$, 
\eqn\eBiv{
\zm (u) = \inv{2u} 
\( 4 (D\eh ) (u) + \xm (u) (\Ch_- + 2) - \xp (u) \Ch_+ \) . } 
Finally, using the expression for $D\eh$ from \EDe, one obtains
the first equation in \eBii. 

The relations \eBii\ can be used to prove the constraint
\EDh\ amoung the $C$-potentials. Namely,
using the definition \eBi\  for $\zm$, and the expression
\eBii\ for $\zm$ in terms of $\xpm$, one has 
\eqn\eBv{\eqalign{
2 \vev{e, D \ehp } &= - \vev{e, \inv{2u} (\Ch_- \xm - \Ch_+ \xp ) } \cr
&= - \inv{2} \( \Ch_- \vev{e' , \xm } - \Ch_+ \vev{e' , \xp } \) \cr
&= - \inv{2} \( \Ch_- (C_- +2) - \Ch_+ C_+ \) .\cr  }}
On the other hand, 
\eqn\eBvi{\eqalign{
2 \vev{ e, D \ehp } &= 2 \vev{ e, D\ehp }^t = 2 \vev{\ehp , Ee } \cr
&= - ( \Ch_- + 2 ) . \cr}}
Comparing \eBv\ and \eBvi, one obtains the constraint \EDh.

\def\col{\pmatrix{\xp \pm \xm \cr ~ \xph \pm \xmh \cr } } 
\def\rpm{ (1- {\bf R_\pm } ) } 
\def\Bh{\hat{B}} 

\appendix{C}{Proof of the linear system.}

In this appendix, we prove the equations \EDr, \EDs. 
Multiplying \eIIIiv\ by $1\pm \kb$, one obtains 
\eqn\eCi{
2\pmatrix{e\cr \eh \cr} 
= (1\pm \kb ) 
\col . } 
Taking the $z$-derivative of this equation, then multiplying on the left
by $1 - {\bf R_\pm }$, leads to 
\eqn\eCii{
\d_z \col = 2 (1- {\bf R_\pm } ) \pmatrix{\d_z e \cr \d_z \eh \cr } 
\mp (1- {\bf R_\pm } ) \d_z \kb \col . } 
 From \EDjj\ and \EDe, one has 
\eqn\eCiii{
\( (1-{\bf R_\pm} ) \d_z \kb \) (u,v) 
= - \frac{m}{4} 
\pmatrix{ \pm \xm (u) e(v) & \xp (u) \eh (v) \cr 
\xph (u) e(v) & \pm \xmh (u) \eh (v) \cr } . } 
Performing a simple multiplication gives 
\eqn\eCiv{
\rpm \d_z \kb \col 
= - \frac{m}{4} (B_+ \pm B_- \pm 2 ) \pmatrix{\pm \xm \cr \xph \cr} 
- \frac{m}{4} (\Bh_+ \pm \Bh_- \pm 2 ) 
\pmatrix{\xp \cr \pm \xmh \cr } . }

The other term on the RHS of \eCii\ may be simplified to 
\eqn\eCv{
\rpm \pmatrix{\d_z e \cr \d_z \eh \cr} 
= - \frac{m}{2} 
\pmatrix{ (1-F)(ue) \mp D(u\eh ) \cr 
\mp E (ue) + (1-G)(u\eh ) \cr } . } 
Expressions of this kind can all be computed similarly. 
This requires commuting the multiplication by $u$ and the
operators $E,\ D,\ F,\ G$. Consider for example 
\eqn\eCvi{\eqalign{
\( D(u\eh ) \) (v) 
&= \int \frac{dw}{w} D(v,w) (w + v -v ) \eh (w) \cr 
&= - v (D\eh ) (v) + \int \frac{dw}{w} (w+v) D(v,w) \eh (w) . \cr }}
Inserting the explicit expression \EDd\ for $D$ in the second term, 
we obtain 
\eqn\eCvii{
\( D(u\eh ) \) (v) = \frac{v}{2} \xm (v) 
+ \inv{4} \( \xp (v) \Bh_+ - \xm (v) (\Bh_- +2) \) . }
Similarly, 
\eqn\eCviii{\eqalign{
\( E(ue ) \) (v) &= \frac{v}{2} \xmh (v) 
+ \inv{4} \( \xph (v) B_+ - \xmh (v) (B_- +2) \) \cr  
\( (1-F)(ue ) \) (v) &= \frac{v}{2} \xp (v) 
+ \inv{4} \( \xp (v) (B_- +2)  - \xm (v) B_+  \) \cr  
\( (1-G)(u \eh ) \) (v) &= \frac{v}{2} \xph (v) 
+ \inv{4} \( \xph (v) (\Bh_- +2)  - \xmh (v) \Bh_+  \) .\cr }}  
Collecting all terms on the RHS of \eCii, and solving for 
$\d_z \xpm , \ \d_z \xpmh$ yields the $\d_z$ part of the
result \EDr. 

Consider now the $\zb$ derivatives.  By the same reasoning
as above, we begin with \eCii\ but with $\d_z$ replaced
with $\d_\zb$.  One needs 
\eqn\eCix{
\rpm \pmatrix{ \d_\zb e \cr \d_\zb \eh \cr } 
= - \frac{m}{4} \pmatrix{\zp \pm \zm \cr \pm \zmh + \zph \cr } , }
where we have used the definition \eBi. 
As in \eCiv, one also has 
\eqn\eCx{
\rpm \d_\zb \kb \col 
= - \frac{m}{4} (C_+ \pm C_- \pm 2 ) \pmatrix{\pm \zm \cr \zph \cr} 
- \frac{m}{4} (\Ch_+ \pm \Ch_- \pm 2 ) 
\pmatrix{\zp \cr \pm \zmh \cr } . }
Putting all of this together, one finds that the result simplifies
greatly:
\eqn\eCxi{\eqalign{
\d_\zb \xpm (u) &= \frac{m}{4} \( C_\pm \zm (u) + \Ch_\mp \zp (u) \) \cr
\d_\zb \xpmh (u) &= \frac{m}{4} 
\( \Ch_\pm \zmh (u) + C_\mp \zph (u) \)  .\cr}}
Finally, using the expressions \eBii\ for $\zpm , \zpmh$ in terms 
of $\xpm , \xpmh$, one obtains the rest of \EDr, \EDs.

\appendix{D}{The Ising case.}

In this appendix, we provide a new proof of the differential
equation for the Ising correlation functions\eIIxxxii,  
and give more details concerning the properties of the kernels
of the type \EBa. 
These correlators are expressed as the
Fredholm determinant for the scalar kernel \eIIxxxivc~ which
is in the class of integral operators \EBa .
Rather than repeat what was done above for the sine-Gordon 
theory, we present the proof in the case where the 
kernel is replaced by a finite dimensional matrix. 
More specifically, we first cutoff the integrals
by replacing $\int_0^\infty du$ with 
$\int_0^\Lambda du$ for $\Lambda$ finite. 
Consider then discretizing the $u$ variables
into $N$ intervals, where in each interval $u$ takes the
value $u_i$, $i = 1,..,N$ :  
\eqn\contin{
\int_0^\Lambda  du \> f(u)  ~\to ~ 
\Delta u \sum_{i=1}^N f(u_i ) , }
where $u_n = n \Delta u$, and $N\Delta u = \Lambda$. 
The integral operators arise
in the continuum limit where $N \to \infty$, 
$\Delta u \to 0$, with $N\Delta u $ kept fixed;
indeed, this is the way in which the Fredholm theory 
is usually developed.  In the end, we take $\Lambda$ to
infinity. 
We do this in order to illustrate a basic point: 
the above proofs do not rely on being in the continuum limit
but are actually valid for finite $N$. 
Of course,  all of the proofs below can 
be done also  in the continuous case as was done above
for sine-Gordon. 

Therefore, we consider two tau functions,
\eqn\EDDa{
\tau_\pm = \det\(1\pm V\),}
where $V$ is an $N\times N$ matrix with entries,
\eqn\EDDb{
V_{ij} = \frac{e_i\ e_j}{u_i+u_j} }
with $(u_i)$ a set of complex numbers.
The $(e_i)$ are functions depending on the parameters $u_i$
as well as on the coordinates $(z,\zb)$.
As explained above, the actual continuum Fredholm determinant \eIIxxxiv\ 
is obtained by letting N go to infinity, identifying
the $u_i$ as a discretization of the continuum variable $u$,
and replacing the
infinite sums with integrals.  
We assume that the $(z,\zb)$ dependence of $V$ comes only from 
the functions $e_i$ through the following formula~:
\eqn\EDDc{\eqalign{
 \d_z e_i &= -\frac{mu_i}{2} e_i , \cr
 \d_\zb e_i &= -\frac{m}{2u_i} e_i  . \cr}}
It is well known that this implies that $\tau_\pm$
are the tau functions of the $N$ soliton solutions of
the classical sine-Gordon equations.
This is the result that we will rederive using our
technique. The method is similar to the case 
described in the main text but all the computations
are actually simpler since the kernel is a scalar. 
The key point is the group property \eIIIii~ with
the $\Z_2$ graded multiplication law.

We first need the resolvent $R_\pm$ of the matrix $(\pm V)$~:
\eqn\EDDe{
 (1- R_\pm)(1\pm V) = 1.}
Similarly to the integral operators \EBa, the matrices of the form
\eqn\EDDd{
\(V_{(\pm)}\)_{ij} = \frac{\sum_a e^a_i e_{a,j}}{u_i\pm u_j}}
form a group with the $\Z_2$ graded multiplication law 
sketched in eq. \eIIIii.
Therefore, the resolvent is an element of the group.
From the equation \EDDe, it is clear that $R_\pm$ has
components of both type $\[V_{(\pm)}\]$. Namely, we have
\eqn\EDDf{
R_\pm = H \pm F }
where $H$ is matrix of type $\[V_{(-)}\]$ and
$F$ a matrix of type $\[V_{(+)}\]$. 
Since the matrix \EDDb~ is of type $\[V_{(+)}\]$, eq. \EDDe~
is equivalent to~:
\eqn\EDDg{
VF+H=0,\quad VH+F=V.}
These equations can be solved, giving the following
expression for the matrices $F$ and $H$~:
\eqn\EDDh{\eqalign{
H_{ij} &= \frac{f^-_if^+_j-f_i^+f_j^-}{2(u_i-u_j)}\cr
F_{ij} &= \frac{f^+_if^-_j+f_i^-f_j^+}{2(u_i+u_j)}\cr}}
with
\eqn\EDDj{
f^\pm_i = \[(1-R_\pm)e\]_i .}
The proof consists in checking directly the relations \EDDg~ using \EDDj.

\def\ep{e'}
\def\fp{{f' \,}^\pm}
Besides the vector $(e_i)$ and $(f^\pm_i)$, we also 
define vectors $(e'_i)$ and $(\fp_i)$ by~:
\eqn\EDDm{
\fp_i = \[(1-R_\pm)\ep \]_i,\quad \ep_i=\frac{e_i}{u_i}.}
We need to introduce the ``potentials".  They are defined by~:
\eqn\EDDk{
P_\pm = \vev{e,f^\pm},\quad
Q_\pm = \vev{\ep ,f^\pm},}
where the scalar product of two vectors with
components $(a_i)$ and $(b_i)$ is~:
\eqn\EDDl{
\vev{a,b} = \sum_i\ a_i b_i .}
The $\fp_i$ and  $f^\pm_i$ are not independent,
but are related by the potentials. 
 We have~:
\eqn\EDDn{
\fp_i = \inv{u_i}\(1\mp Q_\pm\)\ f^\mp_i.}
The proof goes as follows. By the definition \EDDm~ we have~:
\eqn\EDDo{
\fp_i= \frac{e_i}{u_i}-\[(H\pm F) \ep \]_i.}
Then, using the formula \EDDh~ for $H$ and $F$, we deduce that~:
\eqn\EDDp{\eqalign{
\(H \ep \)_i =\inv{2u_i}\(f^-Q_+-f^+Q_-\) + \inv{u_i}\(He\)_i\cr
\(F \ep \)_i =\inv{2u_i}\(f^-Q_++f^+Q_-\) - \inv{u_i}\(Fe\)_i.\cr}}
Substituting the latter into \EDDo~ proves \EDDn.

The relation \EDDn~ implies that the potentials $Q_\pm$ are
not independent. Indeed, taking the scalar product of $\fp$  
with $(e_i)$ and using $\vev{e, \fp }=\vev{\ep ,f^\pm}=Q_\pm$,
we obtain~:
\eqn\EDDq{
(1-Q_+)(1+Q_-)=1.}

The derivatives of the logarithm of the tau functions
\EDDa~ are given by the potentials. Using 
$\d_z\log\tau_\pm=\pm{\rm tr}\[(1-R_\pm)\d_zV\]$
and \EDDc, we obtain~:
\eqn\EDDr{\eqalign{
\d_z\log\tau_\pm &= \mp\frac{m}{2} \vev{e,f^\pm}\cr
\d_\zb\log\tau_\pm &= \mp\frac{m}{2} \vev{\ep ,\fp }\cr}}

The next step consists in finding the auxiliary linear system.
This is a system of linear differential equations for
the $N$ two-dimensional vectors $\pmatrix{f_i^+\cr f^-_i\cr}$.
Let us first describe the result and then give the proof.
The system is~:
\eqn\EDDt{\eqalign{
\d_z \pmatrix{f_i^+\cr f^-_i\cr} &=-\frac{m}{2}
\pmatrix{ -P^+-P^- & u_i \cr u_i & P^++P^-\cr}
\pmatrix{f_i^+\cr f^-_i\cr} \cr
~&~ \cr
\d_\zb \pmatrix{f_i^+\cr f^-_i\cr} &= -\frac{m}{2 u_i} 
\pmatrix{ 0 & (1-Q^+)^2 \cr
 (1+Q^-)^2 & 0 \cr } \pmatrix{f_i^+\cr f^-_i\cr} . \cr }}
Let us first prove the $\d_z$ equations.
Taking the derivative of \EDDj~ and using the expression \EDDh~ for the
resolvent, we obtain~:
\eqn\EDDu{
\d_z f^\pm_i = \pm\frac{m}{2}P_\pm f^\pm_i -\frac{m}{2}
\[(1-R_\pm)(ue)\]_i .}
Now, the last term in \EDDu~ can be simplified using \EDDh~:
\eqn\EDDv{
\[(1-R_\pm)(ue)\]_i = \mp f^\pm_i P_\mp
+ u_i \[(1-R_\mp)e\]_i .}
Inserting this relation into \EDDu~ gives the $\d_z$ part of \EDDt.
The $\d_\zb$ equations follow from~:
\eqn\EDDw{
\d_\zb f^\pm_i = -\frac{m}{2} \(1\mp Q_\pm\) \fp_i,}
which is a direct consequence of \EDDc~ and of the definition \EDDj.
Expressing $(\fp_i)$ via equation \EDDn~
gives the $\d_\zb$ part of \EDDt.

We now relate the potentials $P_\pm$ to the
$Q_\pm$. The $\d_z$ derivative of $Q_\pm$ is~:
\eqn\EDDaa{
\d_z Q_\pm = \vev{\d_z \ep ,f^\pm} + 
\vev{\ep ,\d_z f^\pm}.}
Using \EDDc~ and the linear system \EDDt, we obtain~:
\eqn\EDDbb{
\d_z Q_\pm =-\frac{m}{2}\(1\mp Q_\pm\)(P_++P_-).}
Solving the algebraic relation \EDDq~ by parametrizing the
potentials $Q_\pm$ by~:
\eqn\EDDcc{
1\mp Q_\pm = \exp(\mp\varphi),}
for some auxiliary function $\varphi$, we deduce that~:
\eqn\EDDdd{
\frac{m}{2}(P_++P_-) = -\d_z\varphi.}

Inserting this parametrization in \EDDt,
one recognizes the Lax pair of the sinh-Gordon model.
Therefore, its zero curvature condition gives~:
\eqn\EDDae{
\d_z\d_\zb\varphi = \frac{m^2}{2} \sinh2\varphi .}
The last step consists now in relating the tau functions
\EDDa~ to the function $\varphi$. Using \EDDr, one 
finds that~:
\eqn\EDDaf{\eqalign{
\d_z\d_\zb\log(\tau_+\tau_-) &= \frac{m^2}{2}\(1-\cosh 2\varphi\)\cr
\d_z\d_\zb\log\(\frac{\tau_+}{\tau_-}\)
&=  \frac{m^2}{2} \sinh2\varphi.\cr}}
These are the equations of motion of the affine
$\hat{sl_2}$ Toda theory \ref\rBB{O. Babelon and L. Bonora, 
Phys. Lett. B244 (1990) 220.}.

Note that since the Ising correlators are $\tau_\pm$ themselves
rather than their product and quotient, the above parametrization
of the Ising correlators is somewhat awkward in comparison with
the sine-Gordon case. This suggests that the differential equations
perhaps have a more natural explanation in the sine-Gordon case.

\listrefs
\end